\documentstyle[11pt,aaspp4]{article}
\received{7/14/97}
\accepted{12/9/97}
\slugcomment{ApJ in press}
\lefthead{Bagoly, \Mesz, Horv\'ath, Bal\'azs \& \Mesz}
\righthead{Principal Component Analysis of 3B Catalog}

\def\p1{\phantom{@}}
\def\p2{\phantom{@@}}

\def\tl{\tableline}
\def\be{\begin{equation}}
\def\ee{\end{equation}}
\def\bc{\begin{center}}
\def\ec{\end{center}}

\begin{document}

\begin{center}
\title{A Principal Component Analysis of the 3B Gamma-Ray Burst Data}
\author{Z. Bagoly,\altaffilmark{1} A. M\'esz\'aros,\altaffilmark{2,3} I. 
Horv\'ath,\altaffilmark{4} L. G. Bal\'azs,\altaffilmark{3} and P. 
M\'esz\'aros\altaffilmark{4}}  
\altaffiltext{1}
{E\"otv\"os University, Laboratory for Information Technology,
M\'uzeum krt. 6-8, Budapest, H-1088, Hungary}
\altaffiltext{2}{ Dpt. Astronomy, Charles University, 150 00
Prague 5, \v{S}v\'edsk\'a 8, Czech Republic}
\altaffiltext{3}{Konkoly Observatory, Budapest, Box 67, H-1525, Hungary}
\altaffiltext{4}{Dpt. Astronomy \& Astrophysics, Pennsylvania State 
University, 525 Davey Lab. University Park, PA 16802}
\today
\end{center}

\begin{abstract}
We have carried out a principal component analysis for 625 gamma-ray 
bursts in the BATSE 3B catalog for which non-zero values exist for the 
nine measured variables. This shows that only two out of the three basic 
quantities of duration, peak flux and fluence are independent, even if
this relation is strongly affected by instrumental effects, and these 
two account for 91.6\% of the total information content.
The next most important variable is the fluence in the fourth energy 
channel (at energies above 320 keV). This has a larger variance and is less 
correlated with the fluences in the remaining three channels than the latter 
correlate among themselves. Thus a separate consideration of the fourth channel,
and increased attention on the related hardness ratio $H43$ appears useful
for future studies. The analysis gives the weights for the individual
measurements needed to define a single duration, peak flux and fluence.
It also shows that, in logarithmic variables, the 
hardness ratio $H32$ is significantly correlated with peak flux, while 
$H43$ is significantly anticorrelated with peak flux. The principal 
component analysis provides a potentially useful tool for estimating the 
improvement in information content to be achieved by considering alternative 
variables or performing various corrections on available measurements.

\end{abstract}

\keywords{gamma rays: bursts -- methods: statistical -- methods: data analysis} 

\section{ Introduction}

Extensive data bases on Gamma-Ray Burst (GRB) properties such as the BATSE 
3B catalogue (\cite{me96}) contain a wealth of statistical information. 
However, to translate that into useful knowledge about the physics of GRB 
requires a significant amount of interpretational effort. In its simplest 
form, the 3B archive, and foreseeably also its later updates, contains 9 
entries per event, including several different definitions of three basic 
physical measurements, the duration, fluence and peak flux.
One of the principal questions that must be asked is how many of these
entries are truly important.  Another question is, which subset or combination
of these nine quantities contains the maximum amount of non-redundant 
information. In other words, what is the number of significant physical 
quantities responsible for the observed variables. Numerous analyses of the 
3B catalogue and its predecessors have been made (e.g. \cite{kou93}; 
\cite{lgs93}; \cite{mit94}; \cite{nor94}; \cite{bri95}; \cite{nor95}; 
\cite{low95}; \cite{emh95}), using different techniques. These papers have 
looked at various statistical properties of the bursts within a broader 
context. However, none of them appear to have investigated the above questions
with a method specifically designed to answer them. In this paper we address 
these questions specifically, within the framework of the Principal Component 
Analysis (PCA), which is particularly suited for this task. The PCA is a well 
known statistical method (e.g. \cite{mor67}, \cite{jol86}, \cite{mur87}) 
that has wide applications in engineering, artificial intelligence, geophysics,
biophysics, and also in some areas of astronomy, e.g. 
\cite{con95}; \cite{bal96}. However, to our knowledge this method has not so 
far been used to any significant extent in the field of GRB (a partial 
exception is \cite{mfb97}). We show that the PCA can provide useful insights 
into the statistical properties of the variables in the 3B catalogue, 
which could simplify the investigation of the physical nature of GRB.

Broadly speaking, there are three main benefits to be gained from an 
analysis using the principal components of a data set. First, without 
any essential loss of information, it allows one to reduce the total 
number $n$ of observed variables to a lower value $m<n$, which can be 
considered to be statistically uncorrelated with each other (i.e. one can 
obtain a smaller number $m$ of ``highly significant" variables). Second, 
it identifies among the remaining $(n-m)$ variables a smaller number that 
may contain some degree of further information. Finally, it allows one to 
obtain, in some sense, information about the character of parameters of the 
sources themselves (e.g.  \cite{con95}; \cite{bal96}). 

The nine entries of the 3B database for each GRB consist of
two durations, $T_{50}$, $T_{90}$, which contain 50\% and 90\% 
of the burst counts, respectively (\cite{kou93}, \cite{kos96}); four 
fluences (time-integrated energy fluxes) $F1,~F2,~F3,~F4$, defined over different energy channels; and three measures of the peak flux (each summed 
over the four energy channels), measured over three different resolution 
time scales (64 ms, 256 ms and 1024 ms).  Thus the initial number of 
variables for the PCA is $n=9$. There is, of course,
some incompleteness in the catalogue, in that not all 9 quantities
are available for all the GRB in the catalogue. There are several possible
ways to deal with this problem in statistical studies 
(e.g. \cite{jol86}; 
p. 219). However, here we will not address the incompleteness issue, choosing
instead to use a subset of GRB for which all nine entries are non-zero.
There are 625 such GRBs in the 3B catalogue, and the PCA is done here on
these. In \S 2 a PCA is done for the subspace of the four fluences only,
to address the question of their independence, and to probe the information
contained in related quantities such as hardness ratios. In \S 3  we perform 
a PCA of the full $n=9$ variables in the catalogue. In \S 4 we discuss
and summarize the results. In Appendix A a similar statistical
method called Factor Analysis is also shortly presented, which
completes the results of \S 3. 

\section{ Principal Component Analysis of the Four Fluences}

A PCA for the fluences in four channels is straightforward, using the method 
of the correlation matrix described in Chapt.3 of \cite{jol86}. A PCA by 
means of the correlation matrix is the default option of a number of 
statistical packages. The motivation for using correlations instead of 
covariances is based on the fact that the observed variables might have very 
different scales. The correlation uses variables normalized with standard 
deviations and mean values, thus overcoming the scale disparity problem.
This is important for GRB, which have durations ranging over six orders of 
magnitude. Also, throughout this paper, we will deal with the logarithms of 
the  quantities involved. This leads to a linear functional dependence 
between two quantities in the case when one quantity is proportional to 
an arbitrary power of another quantity. This is a reasonable 
assumption in our case since a number of models of GRB-s
predict power law relationship between the observable quantities; cf. 
\cite{nem94}; \cite{pac95}.  For this reason, the use of
logarithms seems to be a justified  procedure. (We note, however,
that we have also done the entire analysis directly on the quantities
themselves, as opposed to their logarithms, and the basic conclusions remain
unchanged). The correlation matrix for the logarithms of the four fluences 
is immediately calculable from the BATSE 3B catalogue, and is presented in 
Table 1.
% TABLE 1 about here

The correlation matrix is symmetric, and is written as usual in a
``triangle form" (see, e.g. \cite{jol86}; p.34). The values in Table 1
are straightforward: for example, 0.97 in the first row and second column 
is the correlation coefficient between the logarithms of the fluences in the 
first and second channels; etc. The correlation coefficients are calculated 
with the classical Pearson's formula (Chapter 14.5. of \cite{pre92}),
the diagonal matrix elements being always unity 
(\cite{jol86}; Chapt.3).
The values in Table 1 indicate that there are extremely strong correlations 
among the fluences of first three channels. On the other hand, the fourth 
channel is obviously less correlated with the remaining ones, although 
the degree of correlation is still significant. In Table 1 and in the rest
of the Tables with correlation matrices in this paper, the symbol $\ddagger$ 
indicates a $\geq 99.9\%$ probability that the two quantities considered 
are correlated, while the symbol $\dagger$ indicates a correlation probability 
between 99\% and 99.9\%. (The probability of the existence of correlation is 
calculated using equation 14.5.2 from \cite{pre92}.)

%BEGINNING OF CHANGE 
The four variables $\lg Fi$ constitute a 4-dimensional vector space, 
and the principal components are, in essence, the eigenvectors of the 
correlation matrix in this space, i.e. they determine  orthogonal directions. 
Usually the principal components are also unit vectors (i.e. the sum of the 
squares of the four coefficients is 1), but one can multiply them with an 
arbitrary non-zero constant without any loss of generality.
The relative weights of the four variables are proportional to the 
coefficients, e.g. if the coefficients are equal (in 4 dimensions they
are 0.5), then the four variables are equally important (they have the
same weight). 
To calculate the principal components (hereafter PCs) one may use, e.g., 
the singular value decomposition algorithm (a numerical routine is available in 
\cite{pre92}) or the iteration method described in Chapt. 7.4 of \cite{mor67}.
The results are shown in Table 2.

% TABLE 2 about here

From Table 2 (first line), the first PC is given by the following linear 
combination of the four basic variables used here:
$0.51 \lg F1\; +\; 0.52 \lg F2\; + \; 0.52 \lg F3\; +\; 0.43 \lg F4$. 
This first PC accounts for $\sim 86\%$ of the total statistical information
(which is 100 \% when the four PCs are taken into account).
One can see that the first PC 
is a unit vector where the weights of the four $\lg Fi$ are almost identical. 
However, the remaining three principal components (the next three lines of 
Table 2) are more complicated combinations of the four $\lg Fi$. For 
instance, the second PC accounts for $\sim 12\%$ of the total statistical
information content in this space, and is given approximately by 
$-\lg F1 - \lg F2 + 2 \lg F4$. This second PC is dominated by $\lg F4$, due
to the much larger weight given to it.
We also see that, in essence, the relative importance of the second PC comes
from the fact that $F4$ does not correlate as strongly with the remaining
three fluences as these do among themselves. Another way to look at the 
second PC is as a quantity involving hardness ratios related to $F4$, i.e.
$-\lg F1 - \lg F2 + 2 \lg F4$ = $\lg F4^2/(\lg (F1 F2))$ = $\lg (H41 H42)$, 
where $Hij = Fi/Fj$ ($i,j = 1,2,3,4$), which is a hardness ratio. 
The hardness ratio $H32$ is more generally used than the other simple ratios 
such as $H21$, etc.  in discussions of the GRB data (e.g. \cite{kou93}).
For completeness, it is necessary to consider six different hardnesses,
of which only three are independent (say $H43, H32, H21$). The remaining three 
hardnesses are obtainable from them ($H42 = H43\;H32$; $H41 =
H42\;H21$; $H31 = H32 \;H21$). We see that the product of two hardnesses
is a PC; i.e. it does not correlate with the remaining three PCs.

We note that the BATSE 3B fluences have associated errors, which 
are listed in the catalog (\cite{me96}). They are calculated by the BATSE
group taking into account both systematical and statistical effects.
The sizes of these errors are sometimes large, and the size of the errors 
can be quite different.   
In order to determine the impact of these errors on our analysis 
we used Monte Carlo simulations. For every burst we obtained new 
$F1, F2, F3, F4$ fluences, chosen randomly out of a distribution around 
the original 3B values, the size of these distributions being determined 
by the specific 3B errors listed for the burst considered. We generated 
100 such new data sets and then repeated the whole data analysis procedure 
for each sample. These simulations show that neither the correlations in 
Table 1 nor the PCs in Table 2 change by more than 0.3\% for different 
realizations of the errors (see also Appendix B). The reason for
this is that the spread of the fluences around the mean (the standard deviation
of the $Fi$) is much larger than either the mean error or the standard
deviation of the errors in the $Fi$. (For some individual bursts, but by no 
means all, the errors can be of order the fluence, but this is not the 
case for the averages).  This means 
that the impact of the listed 3B errors on our PCA calculations is small. 
(We also note here that Monte Carlo simulations of the data errors were done 
for all the other correlation matrices and PCAs presented in this article. 
In all cases the changes implied by such simulations were not in excess of 
$1\%$. Hence, the impact of the errors seems to be unimportant throughout, 
at least for the present purposes).
%END OF CHANGE

In the present case we have $n=4$ variables in our vector space, and we ask 
what is the number $m \leq n$ which are highly significant. There are 
different criteria for the 
definition of $m$ (see \cite{jol86}; Chapt. 6.1.2). For example, in accordance
with Jolliffe's rule (\cite{jol72}), $m$ is given by the number of principal
components which explain more than $70/n$ percent of the variations. This
cut-off level is 17.7\% in our case, which is clearly not fulfilled for the
second PC. Hence $m=1$ according to this criterion. 
Nonetheless, while clearly the third and fourth PCs can safely be assumed to
be ``fully unimportant", the second PC, which accounts for $\simeq 12 \%$ of 
the variation, cannot be considered to be "negligible". 

Summarizing this section's results, one can say that, as a rough approximation,
most of the information in the four logarithmic fluences is contained in the 
first PC, which is the sum of the logarithms of the fluences in the four 
channels. In a second, more precise approximation, the information can be 
represented by two important quantities, or PCs, which are a) the sum of the 
logarithms of the four fluences, and b) the logarithm of the fluence in the 
fourth channel.
%BEGINNING OF CHANGE
We note that the same is true (with similar levels of probability of
correlation) when the PCA is done for the 3B fluences themselves, rather
than than their logarithms. In this case the first PC is in essence the 
total fluence $F=F1+ F2+F3+F4$, while the second meaningful PC is $F4$. 
%END OF CHANGE

\section{Principal Component Analysis with Nine Variables}

In this section we do a PCA on the 625 bursts in the 3B catalogue for which 
all nine quantities ($T_{50},\; T_{90}$, 4 fluences and the peak fluxes on 
%BEGIN CHANGE
three triggers) are non-zero. Again we use as our basic vector space the 
logarithms of the quantities, rather than the quantities themselves. 
%END CHANGE
The correlation matrix is given in Table 3.
The PCs of this 9-dimensional space, and the percent of
variation involved in each of them are shown in Table 4.

% TABLE 3 about here
% TABLE 4 about here

From Tables 
3 and 4 we see that the first PC is again roughly given by the sum 
of all nine logarithmic quantities (durations, fluences, peak fluxes), with 
some extra weight on the first three fluences. Because of the different 
dimensions involved, it has only a formal meaning. The second PC, accounting 
for 26.5\% of the variation is clearly important, this value being far above 
Jolliffe's 70/9 = 7.8\% cut-off level. This PC is roughly given by the formal 
difference of the logarithmic peak fluxes and durations (the sum would be 
along the direction of the fluence, but the difference is orthogonal, as 
expected for different PCs).  This means that the duration, peak flux and 
total fluence are undoubtedly important quantities, but only two of them are 
independent. The third PC, practically defined by $F4$ alone, accounts for 
5.1\% of the variation and is already below Jolliffe's level. Hence $m=2$.
However, because the third PC is just below Jolliffe's level, it might still
be of some importance.  (This is examined in more detail in Appendix A).
The fourth PC, with its 1.5 \%, is far below Jolliffe's limit, and its
importance should be even more questionable.

It is essential, in discussing the PCA of this Section, to consider 
also the importance of some instrumental effects. As seen from Table 3,
it is interesting that the duration is anticorrelated with the peak flux on 
64 ms, it is non-correlated on 256 ms, and it is positively correlated 
on 1024 ms. In fact, there is a controversy among different authors 
in this respect, since \cite{nor94} and \cite{nor95} see an anticorrelation
on 256 ms, but \cite{mit94} does not. These are problematic questions, and 
we think that instrumental effects should play a significant role in this 
behavior. For example, there are strong grounds to argue that the correlation 
between the peak flux on 1024 ms and the duration should have an instrumental 
origin (\cite{lep96}; \cite{lep97}). The same situation should occur also for 
the large correlation between fluences and durations (\cite{lep97}). 
Fortunately, these ambiguities do not change the conclusion that there 
are two important PC's, and that the duration itself is an independent 
quantity. 

% TABLE 5 about here

%BEGIN CHANGE
One can, of course, also use other quantities as the original variables of
the vector space. For instance, we have also done the same analysis with the 
quantities themselves, rather than the logarithms, and the results are 
essentially similar. There appears to be no general rule for preferring 
logarithms over the quantities themselves. Taking logarithms has some numerical 
advantages when dealing with quantities that vary by many orders of magnitude.
%END CHANGE
Another possibility is, instead of using the four fluences (or their 
logarithms), to take their ratios (hardnesses). This is possible because 
there is a one-to-one correspondence between the four fluences and the four 
new variables defined by the total fluence and three independent hardnesses. 
(One has then $F = F1+F2+F3+F4$, $H21 = F2/F1$, $H32 = F3/F2$, $H43 = F4/F3$,
and these four new variables are defined unambiguously by the four original 
ones. The inverse is $F1 = F/(H21 + H21 H32 + H21 H32 + H43 H32 H21)$,
the calculation of the remaining three fluences being obvious). It is
interesting to calculate the correlation matrix among these new quantities
(again taking logarithms). These correlation coefficients are presented in 
Table 5.

The correlation matrix of Table 5 shows several things. 
First, it is clear that all the hardnesses are anticorrelated 
(to $\geq 99.9\%$ significance) with the durations.
This fact is, of course, not new, because e.g. in \cite{kou93} the
same anticorrelation between $H32$ and $T_{90}$ is also presented for 222 
GRBs. Second, it seems also that the hardnesses do not correlate with 
the total fluence. This result is in principle expected from the discussion 
in \S 2. (In the PCA for the subspace of the four fluences by themselves we 
obtained that the total fluence $F$ and the product of $H41$ and $H42$ are 
PCs, and hence they should not correlate. This suggests that also the 
individual hardnesses themselves should not correlate strongly with total 
fluence.) Also, the hardness ratio $H32$ is significantly correlated 
($\geq 99.9\%$) with the peak fluxes $P_{64}$ and $P_{256}$; but 
interestingly, the hardness ratio $H43$ is {\it anti}-correlated with 
the peak flux $P_{1024}$, also with $\geq 99.9\%$ significance.

A computation of the PCs corresponding to Table 5 shows that the first PC
(34\% variation) is dominated by the peak flux, the second PC (30\%) by the 
duration (both with contributions from the fluence), and the third PC (15\%)
is dominated by $H43$.

\section{Discussion and Conclusions}

We have carried out a principal component analysis (PCA) with the
nine variables describing 625 GRBs in the BATSE 3B catalogue. The results 
of this analysis may be summarized as follows.

1. A PCA for the $n=9$ variables identifies a subset of $m=2$ important
variables; i.e. two principal components (PCs) are unambiguously important, 
when Jolliffe's criterion is applied. These are constructed out of the 
fluence, peak flux and duration, implying that only two of the three are 
independent.  This means that in the roughest approximation, it 
is enough to consider, e.g., a total fluence and a duration, and these 
two represent 91.6\% of the information content in the 3B catalogue. 
While in an ideal case the above dependence is obvious from the definition of 
these quantities, it is not clear that it should continue to hold for sources 
in a flux limited sample with complicated light curves, located at increasing 
distances and subject to complicated detection biases. However our analysis
shows that this result is valid for the 3B sample of sources, even if the 
relation between the fluence and the duration (or peak flux) is heavily 
influenced by instrumental effects.

2. From the remaining PCs, the third one is definitely non-negligible,
although strictly it is already below Jolliffe's cut-off level. This
third PC is roughly identical to the fluence in the fourth (highest energy)
channel, $F4$. This means that in a finer approximation one could take, e.g.,
duration, total fluence and $F4$ as variables containing significant 
information. The fact that $F4$ has a different behavior than the remaining 
three fluences was also confirmed by a PCA of the subset of four fluences 
alone. This different behavior of the fourth channel also is manifested by 
the fact that the hardness ratio $H43$ has a much bigger variance than the 
commonly used $H32$. The first three PCs account for 96.7\% of the total
information available in the 3B catalogue.

3. A fourth PC is defined, in essence, by the ratio of fluence and peak
flux. Thus in an even finer approximation, one would in addition to the
above three PCs also consider the information provided by considering
separately the fluence and the peak flux. However, the latter does not 
add much new information (only 1.5\% more, see Table 4). In other words, 
in the finest approximation one can take, e.g., the total fluence, duration,
$F4$ and peak flux, and this contains 98.2\% of the total information
content of the 3B catalogue with nine entries per burst.

%BEGIN CHANGE
4. The most meaningful single value of the duration, fluence and peak flux 
for each event which can be constructed from the nine entries in the 3B
catalog is obtained by using the weights given in the first line of Table 4
for that physical variable. (E.g., a duration would be defined by using
relative weights of 0.29 and 0.31 for $T_{50}$ and $T_{90}$, with an
appropriate normalization, etc.)
%END CHANGE

The above statements are purely statistical, and deliberately omit any
extra information concerning the operational way in which quantities are
measured, or theoretical models of what the data may mean. 
The fact that the fluence and durations are the most important quantities,
and that a hardness and the peak flux are also useful, is agreed upon
at a qualitative level by most people in the field. However, what is new 
here is the more rigorous quantification of the level of importance that
can be assigned by the PCA method to each quantum of information, and the 
corresponding ordering or prioritization of the different quantities that 
this affords. This quantification also allows one, for instance, to decide
whether some alternative definitions of the basic quantities contain more
information than others. 

      From a statistical information viewpoint, there appears to be no 
significant preference between, e.g., the durations $T_{90}$ and 
$T_{50}$; and on the same basis, a choice between the peak fluxes on 
the trigger time scales would appear to be approximately indifferent. Of 
course, additional instrumental, operational or physical model considerations 
would serve to refine such choices, depending on what one is after or
what hypothesis one wants to test. For example, in \cite{che97} remarkable 
conclusions are drawn from the differences between the peak fluxes on the 
64 ms and 1024 ms triggers. 

Some of the results are unexpected. For instance, the analysis indicates
that an allowable approximation would be to combine (add) the fluences in
the first three channels, and consider them in conjunction with the fluence
$F4$ in the fourth channel as the basis vectors for the fluence space.
This singling out of $F4$ based on its (statistical) information content 
appears to be new. As is known, the hardness ratio $H32$ is most often used 
in statistical analyses of the BATSE data (or sometimes $H21$). However, 
$H32$ appears to have a significantly smaller variance than $H43$. 
It seems paradoxical, from the information content viewpoint, to concentrate
attention on a variable of relatively small variance, while generally 
ignoring other variables which have a much greater variance, namely $H43$. 
Of course, a careful consideration is required of whether the greater 
variance of $H43$ is due to greater photon noise or other larger errors
in determining $F4$, or to physically interesting facts (such as,
for example, the spectral break occurring in the fourth channel, or an
additional steepening of the spectral index occurring there, etc.).
%BEGIN CHANGE
A concern here is that (as we understand it) the BATSE channel 4 fluence 
listed in the 3B catalog is obtained from a fit of a lognormal function 
across all four channels, and in the range above 300 keV there is only one
data point to anchor the extrapolation of this model shape (and the 3B usage
of fluences in ergs further accentuates the uncertainty of this extrapolation).
One way to address the importance of errors in $F4$ is by noting the fact 
that, although the  error of $F4$ is much higher than that of 
the remaining three channels, the variance of the logarithms of the $Fi$ 
(discussed in this work) turned out to be roughly the same. The higher 
variance in $H43$ is therefore probably explained by the significantly 
lower correlation of $F4$ with the other three channels (see Table 1). 
We note here again that, as shown by the Monte Carlo simulations, a 
consideration of the errors listed in the 3B catalog does not change the 
singling out of $F4$, and its smaller degree of correlation to the other three 
(see also Appendix B).  Since the first three channels correlate 
pretty well among themselves
they may be explained by the same PCs.  On the contrary, due to its lower 
correlation to the other three, the fourth channel indicates the need for 
a further PC to fully account for it. Hence the fourth channel indicates 
some sort of additional information which is not contained in the other three. 
We strongly suspect that this additional information is of a physical nature. 
(E.g., it could be related to the difference between HE and NHE pulses and 
bursts, Pendleton et al., 1997).
%END OF CHANGE 

Another reason why $H43$ may be of interest is that it shows a significant
anticorrelation with the peak flux, whereas $H32$ shows a correlation 
(Table 5). The PCA analysis presented here, in any case, suggests that 
more information may be available from a careful analysis of quantities
involving the fourth channel than has been previously realized.

Finally, we note that the PCA offers a simple method to estimate the
degree of improvement in information content that is potentially achievable 
by performing different manipulations of the data set beyond what is
made in the 3B catalog or its future incarnations. For instance,
in designing new analyses which involve various corrections (instrumental
or otherwise) to the data, one can use the PCA to measure quantitatively
the increase, if any, in the amount of information available from new 
definition of the variables to be studied. The PCA is therefore a tool
of significant potential usefulness in planning data analysis strategies.

\acknowledgments
This research was partly supported by
NASA NAG5-2362, NAG5-2857 (P.M., I.H.), 
Domus Hungarica Scientiarium Artiumque grant (A.M.), 
OTKA grants F14324, T14304, F26666 (I.H.), T024027 
(L.G.B.) and Sz\'echenyi Foundation (I.H.). A.M. acknowledges the kind 
hospitality of Konkoly Observatory, and we are grateful to E.D. Feigelson,
E.E. Fenimore, G. Pendleton and a referee for useful comments and discussion.

\appendix 
\section{3B Factor Analysis with Nine Variables}
\label{app:factor}

The purpose of this Appendix is to confirm the results of \S 3  
by the use of the Factor Analysis (FA). The FA is a statistical method 
which is closely related but not fully identical to the PCA. (The comparison 
of these two methods is discussed in detail, e.g., in Chapter 7 of 
\cite{jol86}). In essence, the major distinction between the PCA and the FA 
comes from the fact that the FA immediately assumes that the observed $n$ 
variables are linear combinations of only $m < n$ variables, which contain
the basic information in the data. Then the fulfillment of this assumption 
is tested. A possible way to test this is the following: if $m$ new variables 
are enough, then using only them one can reproduce the correlation matrix 
from these $m$ quantities themselves with a high accuracy. From the level of 
this accuracy one may deduce the 
correctness of the assumption (for details 
see \cite{jol86}; \cite{KS76}).

In our case this statistical method may be useful, because it gives
a further independent criterion besides that given by Jolliffe. In \S 3 for 
$n=9$ one obtained $m=2$ from Jolliffe's criterion. Nevertheless, as noted 
there, the third PC was just barely below the threshold of significance,
and therefore one asks whether it should be considered or not.
In \S 3 the correlation matrix (Table 3), and the results of a PCA
(Table 4) are presented, and $m=2$ is deduced. Taking $m=2$ one may
try  to reproduce the correlation matrix using only these two PCs.
The results of this procedure are shown in Table 6. 
% TABLE 6 about here
The "reproduced correlation matrix" is presented in the lower
left triangle. For comparison, in the upper right triangle the differences
("residuals") between the observed correlations and the reproduced 
correlations are presented. One sees straightforwardly that the largest
differences arise when the correlations involve $F4$. On the diagonal 
(marked with asterisks) one should have values close to unity if the PCs 
considered explain well the corresponding observed variable. Clearly,
the departures from unity are again much larger for $F4$ than for any other 
quantity. Performing the test using the quantity given by the equation 
43.132 of \cite{KS76} we came to the conclusion  that the assumption of 
$m=2$ does probably not fully account for the quantity $F4$, and therefore
a third PC is probably required. 

%\appendix
\section{Correlation Coefficients and Errors}
\label{app:corr}

A measurement of two statistical variables $x$ and $y$ subject to measurement 
errors $e_x$ and $e_y$ results in specific values 
$$ \bar{x} = x + e_x, \;\;\; \bar{y} = y + e_y$$ 
where $x$ and $y$ are the ideal values in the absence of errors.
The covariance of the two variables is 
$$ <\bar{x}, \bar{y}> = <x,y> + <x, e_y> + <e_x,y> + <e_x, e_y> $$
and unless the errors are of an unusual type, in the RHS the only nonzero
term is $<x,y>$. This is beacuse one expects the $x$ ($y$) and the 
$e_y$ ($e_x$) to be independent, so their covariances vanish, leading to
$<\bar{x}, \bar{y}> = <x,y>$. Thus, the effect of errors is expected to be 
negligible in the covariance.

The correlation coefficent in the absence of errors is
$$r_{xy} = \frac{<x,y>}{\sigma_x \sigma_y}$$ 
where $\sigma^2_x = <x,x>$ and $\sigma^2_y =<y,y>$. In the presence of
errors, however, we have 
$$\bar{r}_{xy} = \frac{<x,y>}{\sigma_{\bar{x}} \sigma_{\bar{y}}}$$ 
where we use the fact that in the numerator there is no change, and
$$\sigma^2_{\bar{x}} = <x,x> + 2 <x, e_x> + <e_x, e_x>$$
While, as before, one expects that $<x, e_x> = 0$, the last term
$<e_x, e_x>$ is nonzero, and the same is true for the  $y$ variable.
However the order of magnitude of this last nonzero term (the "dispersion
of the error") is very different from that of the dispersion of $x$, and 
in fact, $<e_x,e_x> \ll <x,x>$. 
It also follows that the measured correlations are generally smaller than the
ideal (error-free) correlations
($\mid \bar{r}_{xy} \mid \leq \mid r_{xy} \mid$).

Concretely, if $x$ is $F3$ and $y$ is $F4$, then the fact that $<e_x,e_x> 
\ll <x,x>$ and $<e_y,e_y> \ll <y,y>$ can be verified directly from the 3B
data base. The catalog shows that, while in some cases $e_y \simeq y$, 
the domain of the $y$ (i.e. the spread of the $F4$) is much greater than 
the domains of the corresponding $e_y$ (the spread of the $e_{F4}$). 
In particular, for channels 1, 2, 3, 4 the standard deviation of the fluences
around the mean and the mean error are given by the following pairs
($\sigma_F \mid {\bar e}$) : $2.18 \mid .02;~ 2.22 \mid .02;~ 8.88\mid .06;~ 
29.93\mid .64$ (and the corresponding standard deviations of the errors around
the mean error are also small, $.02,~.02,~.07,~.74$).
Thus the fact that the errors are sometimes comparable to the measured
value ($ \mid e_y \mid \simeq y$, in a fraction of the cases) does not 
mean that the correlation coefficients vary by much. The effect of the errors 
on the correlations is nonzero, but from the above it is expected to be small, 
and there should be no relationship between the magnitude of the difference 
between $r_{xy}$ and $\bar{r}_{xy}$ and the number of GRB measured.

This is verified by the Monte Carlo simulations, which used the errors 
listed in the 3B catalog for every burst to determine a distribution out 
of which new values of $F1, F2, F3, F4$ were chosen at random.
This was done for every burst in the sample, and repeated 100 times, to
calculate new correlation matrices and new PCs.  The results varied at most by
$\sim 0.3\%$. This shows that the fact that there are large errors in some
bursts does not affect the conclusions about the correlation or
lack of correlation between $F4$ and the other variables.  Similar Monte Carlo
simulations were done for the nine variable case, and for all correlations
and PCs discussed in this paper. The results never varied 
by more than a percent.

% TABLES
%\clearpage
\begin{table*}
\begin{center} 
\begin{tabular}{rrrrr}
\tl\tl
$lgF1$ & $lgF2$ & $lgF3$ & $lgF4$ & \cr \tl
1\phantom{@@} &  0.97$\ddagger$ &  0.90$\ddagger$  &  0.62$\ddagger$ & $lgF1$ \cr
   & 1\phantom{@@}  &  0.94$\ddagger$  &  0.65$\ddagger$ & $lgF2$ \cr
   &    & 1\phantom{@@}   &  0.75$\ddagger$ & $lgF3$ \cr
 & &  &   1\phantom{@@} & $lgF4$ \cr 
\tl
\end{tabular} 
\end{center}

\tablenum{1}
\caption{Correlation matrix between the logarithms of 
the fluences in the four channels for 625 GRBs in the 3B catalogue.
The $\ddagger $ symbol (usually written as $**$ in the statistical literature),
indicates that the quantities in that row and column are correlated
with a probability $\geq 99.9\%$.}

\end{table*}

%\clearpage
\begin{table*}
\begin{center} 
\begin{tabular}{rrrrr}\tl\tl
\%  & $lgF1$ & $lgF2$ & $lgF3$ & $lgF4$ \cr \tl
85.75  & 0.51 &  0.52 &  0.52&  0.43\cr
11.75 & -0.37 & -0.32 & -0.04 &  0.87\cr
2.00  &   0.56 & 0.06 &  -0.78 &  0.23\cr
0.50  & 0.53 & -0.79 &  0.31 & -0.05\cr
\tl
\end{tabular}
\end{center}

\tablenum{2}
\caption{Principal components of the logarithms of the four
fluences. In the first column the percentage of total variation 
along that particular PC is given.}
\end{table*}

%\clearpage

\begin{table*}
\begin{center}
\begin{tabular}{rrrrrrrrrl}\tl\tl
$lgT_{50}$ & $lgT_{90}$ & $lgF1$ & $lgF2$ & $lgF3$ & $lgF4$ & $lgP_{64}$
 & $lgP_{256}$ &  $lgP_{1024}$ & \cr \tl
1\phantom{@@}&0.97$\ddagger$ &0.80$\ddagger$ &0.78$\ddagger$ &0.71$\ddagger$
&0.44$\ddagger$  &-0.16$\ddagger$ & -0.01\phantom{$\dagger$} &0.29$\ddagger$  & 
$lgT_{50~~}$  \cr
&1\phantom{@@}&0.83$\ddagger$ &0.81$\ddagger$ &0.74$\ddagger$ & 0.48$\ddagger$ 
&-0.11$\dagger$&
                    0.05\phantom{$\dagger$} & 0.35$\ddagger$  & $lgT_{90~~}$  \cr
 && 1\phantom{@@}& 0.97$\ddagger$  & 0.90$\ddagger$ & 0.62$\ddagger$  &
0.30$\ddagger$  & 0.44$\ddagger$  & 0.69$\ddagger$  & $lgF1$  \cr
 &&&        1\phantom{@@} & 0.94$\ddagger$  & 0.65$\ddagger$  & 0.35$\ddagger$  &
0.49$\ddagger$  & 0.73$\ddagger$  & $lgF2~$  \cr
 &&&&        1\phantom{@@}& 0.75$\ddagger$  & 0.47$\ddagger$  & 0.60$\ddagger$  &
0.80$\ddagger$  & $lgF3~$   \cr
 &&&&        &  1\phantom{@@}& 0.46$\ddagger$  & 0.53$\ddagger$  &
0.63$\ddagger$  & $lgF4~$   \cr
 &&&&     &   &1\phantom{@@} & 0.97$\ddagger$  & 0.84$\ddagger$ & $lgP_{64~~}$   \cr
  &&&&   &   &   & 1\phantom{@@} & 0.93$\ddagger$  & $lgP_{256~}$   \cr
  &&&&        &   &   &  & 1\phantom{@@} & $lgP_{1024}$   \cr
\tl
\end{tabular}
\end{center}
\tablenum{3}
\caption{Correlation matrix of the logarithms of the two durations,
four fluences and three peak fluxes for 625 GRBs in the 3B catalogue.
A $\ddagger$ means that the probability of the existence of
correlation (or for a negative sign the existence of anticorrelation)
is greater than 99.9\%, while $\dagger$ means that this probability is
between 99\% and 99.9\%. }
\end{table*}

%\clearpage
\begin{table*}
\begin{center}
\begin{tabular}{rrrrrrrrrr}\tl\tl
\% & $lgT_{50}$ & $lgT_{90}$ & $lgF1$ & $lgF2$ & $lgF3$ & $lgF4$ & $lgP_{64}$ & 
$lgP_{256}$ & $lgP_{1024}$ \cr \tl
64.8 & 0.29 & 0.31 & 0.39 & 0.39 &   0.40  &  0.32 & 0.22 & 0.28 & 0.35 \cr
26.8 & -0.44& -0.41 & -0.16 & -0.13 & -0.04 & 0.07 & 0.53 & 0.47 & 0.30 \cr
5.1 & -0.07 & -0.07 & -0.19 & -0.18 & 0.03 & 0.93 & -0.09 & -0.14 &-0.19\cr
1.5 & -0.48 & -0.44 & 0.53 & 0.42 & 0.11 &   0.04 & -0.20 & -0.24 &-0.11\cr
0.8 & -0.08 & -0.13 & -0.48 & 0.05 & 0.82 & -0.16 & -0.16 & -0.10 &0.03\cr
0.5 & -0.12 & 0.01 & -0.04 & -0.13 & -0.15 & 0.06 & -0.62 & 0.04 &0.75\cr
0.2 & -0.68 & 0.71  & 0.00 & -0.04 & 0.06 & -0.02 &  0.05 & -0.02 &-0.08\cr
0.2 & -0.03 & 0.05 & -0.53 &  0.77 & -0.34 &  0.06 & 0.03 & -0.06 &0.05\cr
0.1 & -0.02 & -0.01 & -0.01 & 0.10 & 0.01 &  0.00 & -0.46 & 0.78 &-0.41\cr
\tl
\end{tabular} 
\end{center}

\tablenum{4}
\caption{Principal components of the logarithms of the two durations,
four fluences and three peak fluxes. The first column gives the percentage 
of total variation in each PC.}
\end{table*}

%\clearpage
\begin{table*}
\begin{center}
\begin{tabular}{rrrrrrrrrl}\tl\tl

$lgT_{50}$ &$lgT_{90}$ & $lgF$ & $lgH21$ & $lgH32$ &  $lgH43$ & $lgP_{64}$
 & $lgP_{256}$ & 
$lgP_{1024}$  \cr \tl
1\phantom{@@} &0.97$\ddagger$ &0.65$\ddagger$ &-0.23$\ddagger$ 
&-0.38$\ddagger$ &-0.33$\ddagger$
 &-0.16$\ddagger$ &-0.01\phantom{$\dagger$}&0.29$\ddagger$ & $lgT_{50}$ \cr
&1\phantom{@@} & 0.68$\ddagger$ &-0.23$\ddagger$ 
&-0.38$\ddagger$  & -0.33$\ddagger$
  & -0.11$\dagger$  & 0.05\phantom{$\dagger$}& 0.35$\ddagger$ & $lgT_{90}$ \cr
&&  1\phantom{@@}  & -0.08\phantom{$\dagger$}& -0.06\phantom{$\dagger$}& 0.03\phantom{$\dagger$}& 0.47$\ddagger$   &
              0.58$\ddagger$  & 0.76$\ddagger$ & $lgF$ \cr
 &&& 1\phantom{@@} & 0.24$\ddagger$ & -0.02\phantom{$\dagger$}& 0.16$\ddagger$ & 0.11$\dagger$ & 
             0.03\phantom{$\dagger$}& $lgH21$ \cr
 &&&& 1\phantom{@@}  & 0.28$\ddagger$ & 0.23$\ddagger$ &  0.17$\ddagger$ &   
0.01\phantom{$\dagger$}& $lgH32$ \cr
 &&&&&  1\phantom{@@}  &   0.02\phantom{$\dagger$}&   
-0.05\phantom{$\dagger$}& -0.18$\ddagger$ & $lgH43$ \cr
 &&&&&&     1\phantom{@@}  &    0.97$\ddagger$ &  0.84$\ddagger$ & $lgP_{64}$ \cr
 &&&&&&&            1\phantom{@@}  &        0.93$\ddagger$   &  $lgP_{256}$ \cr
 &&&&&&&&                                    1\phantom{@@} & $lgP_{1024}$ \cr
\tl
\end{tabular} 
\end{center}

\tablenum{5}
\caption{Correlation matrix for the logarithms of the 
two durations, the total fluence $F=F1+F2+F3+F4$, the hardnesses 
$H21$, $H32$, $H43$ and the peak fluxes for 625 GRBs in the 3B catalogue. 
The meaning of $\ddagger$ and $\dagger$ is given in Table 3.}
\end{table*}

%\clearpage
\begin{table*}
\begin{center}
\begin{tabular}{rrrrrrrrrr}\tl\tl
 $lgT_{50}$ & $lgT_{90}$ & $lgF1$ & $lgF2$ & $lgF3$  & $lgF4$ & $lgP_{64}$
 & $lgP_{256}$ & $lgP_{1024}$ &
 \cr \tl
.955* & .019 & -.024 &-.020 &-.012 & -.030 & .019 & .020 & .009 & $lg T_{50}$\cr

.955 & .958*  &-.021 &-.019 &-.014&  -.031 &  .017 & .019 & .011& $lg T_{90}$\cr

.820 & .848 & .925* &.036 &-.019 &  -.073 &-.000 &-.001 & .005 & $lg F1$ \cr

.803 & .834 & .936 & .950* &.002 &  -.073 &-.001 &-.003 & .005 & $lg F2$ \cr

.722 & .759 & .918 & .939 & .948* & .004 &-.009 &-.011 &-.008 & $lg F3$ \cr

.473 & .509 & .693 & .718 & .745 &.607* &-.041 &-.058 &-.076 & $lg F4$ \cr

-.183 &-.125 & .297 & .352 & .475 & .499 & .969* &.010 &-.009 & $lg P_{64}$ \cr

-.028 & .031 & .440 & .495 & .606 & .589 & .961 & .979* &.014 & $lg P_{256}$\cr

.281 & .336 & .680 & .726 & .805 & .709 & .849 & .916 & .955* & $lg P_{1024}$\cr
\cr \tl
\end{tabular}

\tablenum{6}
\caption{Results of the factor analysis of the nine
quantities used in \S 3. The lower left triangle contains the reproduced 
correlation matrix; the upper right triangle contains the residuals between
the observed correlations and the reproduced correlations; asterisks indicate
the diagonal elements of the reproduced correlation matrix.}
\end{center}
\end{table*}

%\clearpage

\end{document}